\documentclass[prb]{revtex4}

\usepackage{graphicx}
\usepackage{dcolumn}
\usepackage{bm}
\usepackage{amsmath}
\usepackage{amssymb}
\usepackage{latexsym}
\usepackage{epsfig}
\usepackage{amsbsy}
\usepackage{array}
\usepackage{amssymb}
\usepackage{setspace}
\usepackage{bm}
\usepackage{epstopdf}
\usepackage{subfigure}
\usepackage{color}
\usepackage{xcolor}

\def\sint{\ifmmode{- \!\!\!\!\!\! \int}
    \else{\hbox{$- \!\!\!\! \int \ $}}\fi}

\begin{document}

\preprint{Physical Review Letters}

\title{Criterion of quantum synchronization and controllable quantum
synchronization based on optomechanical system}

\author{Wenlin Li, Chong Li\footnote{Corresponding author. E-mail:lichong@dlut.edu.cn} and Heshan Song\footnote{Corresponding author. E-mail:hssong@dlut.edu.cn}}
\affiliation{School of Physics and Optoelectronic Engineering, Dalian University of Technology, Dalian 116024}


\begin{abstract}
We propose a quantitative criterion to determine whether the coupled quantum systems can achieve complete synchronization or phase synchronization in the process of analyzing quantum synchronization. Adopting the criterion, we discuss the quantum synchronization effects between optomechanical systems and find that the error between the systems and the fluctuation of error are sensitive to coupling intensity by calculating the largest Lyapunov exponent of the model and quantum fluctuation, respectively. Through taking the appropriate coupling intensity, we can control quantum synchronization even under different logical relationship between switches. Finally, we simulate the dynamical evolution of the system to verify the quantum synchronization criterion and to show the ability of synchronization control.
\end{abstract}



\maketitle

\section{Introduction}  

In recent years, synchronization effects of two or more interconnected classical systems have aroused comprehensive attention because synchronization phenomena are found widely in nature. For examples, Huygens found that two clocks with different swings at the initial time will appear synchronization with time evolvement; it was also observed that the fireflies glow synchronously and the oscillation of heart cells in human or animal will keep in step with each other. At the same time, synchronization effects exhibit inimitable application potential in many fields, such as the synchronous transmission of information in the Internet, the synchronous transmission and amplification of signals between coupled lasers, the encryption and decryption of signals using chaotic synchronization technology, and so on. To date, the synchronization of classical system has gradually become the investigation focus in the numerous fields. The groundbreaking work on theoretical exploration of classical synchronization is marked by Yamada and Fujisaka who put forward a criterion to judge synchronization behaviors through calculating the Lyapunov exponent of the coupling system and obtained synchronization conditions [1]. After this, Pecora and Carroll found synchronization phenomena in the electronic circuit and designed a circuit scheme of encrypted communications using synchronization techniques, which demonstrates the attractive application prospect of synchronization effects and arouses the intense research interest for synchronization theory and application [2]. Recently, many effective synchronization techniques have been proposed in order to achieve complete or phase synchronization of classical systems [3-7].

Naturally, it is expected to found a similar synchronization phenomenon in quantum systems in order to realize the synchronous transmission of quantum information or states due to its unique advantages of synchronization effects. However, it is difficult to define precisely some concepts which describe synchronization in quantum systems, like ``tracks'' and ``errors". Even some concepts used in classical dynamics are completely unsuitable to be used in quantum dynamics because of remarkable differences of two kinds of systems. Therefore, the relative research of quantum synchronization once is thought as unfeasible. The optomechanical system, as a representative of mesoscopic systems, has attracted widespread attention and systematic discussion recently [8-11,26]. Mesoscopic systems exhibit simultaneously both properties of classical and quantum system under certain conditions because the scale of the system is in-between macro-system and micro-system. So some phenomena, no matter what chaos behaviors and limit cycle in classical kingdom [12-14,23] or quantum entanglement and quantum coherent in quantum domain [15-18,23], have been observed in optomechanical systems, which provides reliable basis to expand synchronization theory from classical to quantum. At 2013, Mari \textit{et al.} extended the classical synchronization concepts to the quantum system [19] and developed quantitative theory of synchronization for continuous variable systems evolving in the quantum regime. And in their work, two different measures quantifying the level of synchronization of coupled continuous variable are also introduced. Whereafter, some progress has been made in interrelated theories and experiments [23-25,27].

However in the general case, the synchronization effects are very sensitive to parameters of the systems, such as driving field, coupled intensity, and so on. Therefore, it is expected further to investigate and obtain a quantitative synchronization criterion in order to determine directly whether the synchronization can be realized. Meanwhile, the synchronization criterion can also be regarded as a necessary and sufficient condition of the synchronization effect, which means that the quantum coupling systems can be adjusted and controlled to satisfy the synchronization criterion and to realize the aim of quantum synchronization. Further, the controllability and practicability of quantum synchronization can be improved.

In this work, we present a general method for discussing synchronization effects in mesoscopic quantum systems. We introduce the first order and second order measurements to describe the expectation value and the fluctuation of error respectively and give the necessary conditions to estimate the presence of quantum synchronization effects. Using this theory, we design a model based on optomechanical system to realize logic control of quantum synchronization. Subsequently, we validate the criterion through the simulation.

This paper is organized as follows: In Sec.\uppercase\expandafter{\romannumeral2}, the classical synchronization theory is briefly introduced. In Sec.\uppercase\expandafter{\romannumeral3}, the processing method of quantum mesoscopic synchronization is described and the quantitative criteria for determining quantum complete synchronization and quantum phase synchronization are proposed. In Sec.\uppercase\expandafter{\romannumeral4}, a controllable quantum synchronization model base on optomechanical system is designed and the phase synchronization effect is discussed. Finally, the summary and the prospects are given in Sec.\uppercase\expandafter{\romannumeral5}.

\section{Classical synchronization theory}  

Considering two classical coupled systems
\begin{equation}
\begin{split}
\partial_{t}x_{1}(t)=F(x_1(t))+U_1(x_1,x_2)\\
\partial_{t}x_{2}(t)=F(x_2(t))+U_2(x_1,x_2)
\end{split}
\end{equation}
where $x_{1}(t)$ and $x_{2}(t)$ are state variables of two systems, $U_{1}$ and $U_{2}$ are couplings between systems, respectively. If the error ${x}_{-}(t)\equiv \left|x_1(t)-x_2(t)\right|\rightarrow 0$ when $t\rightarrow \infty $, the complete synchronization between classical systems is realized. If the phases $\phi_1(t)$ and $\phi_2(t)$ of $x_{1}(t)$ and $x_{2}(t)$ meet ${\phi}_{-}(t)\equiv \left|\phi_1(t)-\phi_2(t)\right|\rightarrow 0$, the phase synchronization between the systems is obtained.

The classical synchronization criteria reported previously are mainly to analyze the stability of the error $x_-(t)$ or phase error  $\phi_-(t)$ and to determine whether 
$x_-(t)$ or $\phi_-(t)$ can converge asymptotically to zero through calculating the largest Lyapunov exponent. In the next section, we will propose a criterion for quantum synchronization based on the classical synchronization theory.

\section{Quantum synchronization criterion}  
In the Heisenberg picture, we use quadrature operators  $q_j(t)$ and  $p_j(t)$ to describe two coupled quantum systems (here $j=1,2$; $[{q}_{j}(t),{p}_{j'}(t)]=i{\delta}_{jj'}$). Hence, the error operators  $q_-(t)$ and  $p_-(t)$ between the systems can be defined as follows
\begin{equation}
\begin{split}
q_-(t)&\equiv [q_1(t)-q_2(t)]/\sqrt{2}\\
p_-(t)&\equiv [p_1(t)-p_2(t)]/\sqrt{2}
\end{split}
\end{equation}

It can be seen from Eq.(2) that the error operators $q_-(t)$ and $p_-(t)$ are physical quantities which describe the differences between the conjugate mechanical quantities of two systems.  However, both $q_-(t)$ and $p_-(t)$ cannot be very small simultaneously due to the Heisenberg uncertainty relation. Therefore, it is required to consider synthetically both values of two error operators in the quantum synchronization measurement. For this reason, Mari \textit{et al.} introduced the following figure of merit based on Eq.(2)
\begin{equation}
S_c(t)={\langle{q_-(t)}^{2}+{p_-(t)}^{2}\rangle}^{-1}
\end{equation}

Eq.(3) is used to gauge the level of quantum complete synchronization and its value is in the range $0<S_c(t)\leq 1$.

Nevertheless, it is also difficult to discuss quantum synchronization by analyzing $S_c(t)$ directly because it is not easy to give a definite criterion for judging the system synchronization or not. On the other hand, the calculation of $S_c(t)$ is also complex in quantum systems. Fortunately, the mean value approximation is acceptable in a mesoscopic system, meaning that we can write a mesoscopic system operator $o(t)$ in the form of $o(t)=\langle o(t)\rangle+\delta o(t)$, here $\langle o(t)\rangle$ is the expectation value of the operator at the moment $t$ and it can be regarded as a description of ``classical properties". $\delta o(t)$ represents the quantum fluctuation of the operator near its expectation value and the quantum effects of the system can be embodied in $\delta o(t)$. For a mesoscopic system, $\delta o(t)$ is small but can not be ignored. Then, on the basis of Eq.(2), the error operators of the systems can be rewritten as follow
\begin{equation}
\begin{split}
q_-(t)&=[(\langle q_1(t)\rangle+\delta q_1(t))-(\langle q_2(t)\rangle+\delta q_2(t))]/\sqrt{2}=\langle q_-(t)\rangle+\delta q_-(t)\\
p_-(t)&=[(\langle p_1(t)\rangle+\delta p_1(t))-(\langle p_2(t)\rangle+\delta p_2(t))]/\sqrt{2}=\langle p_-(t)\rangle+\delta p_-(t)
\end{split}
\end{equation}
where $\delta q_-(t)=[\delta q_1(t)-\delta q_2(t)]/\sqrt{2}$ and  $\delta p_-(t)=[\delta p_1(t)-\delta p_2(t)]/\sqrt{2}$.As discussed above, the quantum effects of the systems are embodied in $\delta o(t)$. Hence, it is only needed to consider $\delta q_-(t)$ and $\delta p_-(t)$ in the discussion of the synchronization deviation influenced by quantum effects and Eq.(3) can be rewritten as
\begin{equation}
S_c'(t)={\langle{\delta q_-(t)}^{2}+{\delta p_-(t)}^{2}\rangle}^{-1}
\end{equation}

It should be noted that $S_c'(t)$ in Eq.(5) is a measurement of the error operator's quantum fluctuations near its expectation value. $S_c'(t)$ will equal $S_c(t)$ and it is a synchronization  measurement only if $\langle q_-(t)\rangle\rightarrow0 $ and $\langle p_-(t)\rangle\rightarrow0 $. Therefore, we define $S_c'(t)$ as a second order measurement to reflect the differences between systems generated by the quantum noise even though the classical synchronization conditions are reached. Correspondingly, $\langle q_-(t)\rangle\rightarrow0 $ and $\langle p_-(t)\rangle\rightarrow0 $ can be defined as a first order measurement of quantum synchronization. Now that $\langle q_-(t)\rangle $ and $\langle p_-(t)\rangle $ satisfy the classical properties, the stability analysis method through calculating the largest Lyapunov exponent of the system can be used to determine whether $\langle q_-(t)\rangle $ and $\langle p_-(t)\rangle $ tend stably to zero. Thus, it is only demanded to control the stability of $\langle q_-(t)\rangle $ and $\langle p_-(t)\rangle $, that is, we can control mesoscopic quantum synchronization instead of analyzing $S_c(t)$ directly.

Similarly, in the discussion of quantum phase synchronization, the phase error operator can be defined as follow
\begin{equation}
\phi_-(t)\equiv [\phi_1(t)-\phi_2(t)]/\sqrt{2}
\end{equation}
here $\phi_j(t)=\arctan[p_j(t)/q_j(t)]$. We introduce $S_p$ as the measurement of the quantum phase synchronization
\begin{equation}
S_p(t)={\dfrac{1}{2}\langle{\phi_-(t)}^{2}\rangle}^{-1}
\end{equation}

Similar to the discussion of complete quantum synchronization, let $\phi_-(t)=\langle \phi_-(t)\rangle+\delta\phi_-(t)$ and rewrite Eq.(7):
\begin{equation}
S_p'(t)={\dfrac{1}{2}\langle{\delta\phi_-(t)}^{2}\rangle}^{-1}
\end{equation}
where $S_p'(t)$ should be a second order measurement as well under $\langle \phi_-(t)\rangle\rightarrow0 $.

In summary, the synchronization effects in mesoscopic quantum systems can be discussed through the following steps:\\
a. Write the operator equations of system's conjugate mechanical quantities in the Heisenberg picture, define the error operators and take them as the form of fluctuations near their expectation value, that is, $o(t)=\langle o(t)\rangle+\delta o(t)$.\\
b. Make stability analysis for $\langle o(t)\rangle $ and calculate the largest Lyapunov exponent of the error equations. If the largest Lyapunov exponent is less than zero, the evolution of $\langle o(t)\rangle $  can tend to zero stably after a certain time, whereas it may be ruleless oscillation. \\
c. If the largest Lyapunov exponent is less than zero, the following work is to discuss the magnitude of the noise ($\delta o(t)$) and to calculate $S_c'(t) $ and $S_p'(t) $ base on Eq.(5) and Eq.(8), respectively. Oppositely, if $S_c'(t) $ and $S_p'(t)$ keeps a constant but not zero, the synchronization between the quantum systems is achieved.

\section{Design of the controlled quantum synchronization model and quantum phase synchronization}  
A controlled quantum synchronization model is designed based on the quantum optomechanical system in order to check the validity of the above-mentioned quantitative criteria. In this model, we can realize quantum synchronization control through different logical relationship of the switches, shown in Fig.1.
\begin{figure}[h!]
\centering
\includegraphics[scale=0.17]{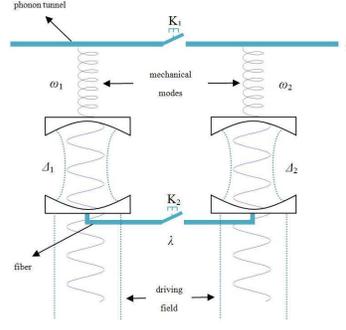}
\caption{Diagrammatic sketch of controlled synchronization model}
\label{threadsVsSync}
\end{figure}

Two coupled optomechanical systems are driven by laser and interact mutually through a phonon tunnel and a fiber which can be controlled by the open or close of the switches $K_1$ and $K_2$. The Hamiltonian of the system can be given directly after a rotating approximation [20, 21] ($ \hbar=1$).
\begin{equation}
\begin{split}
H=\sum_{j=1,2}&[-\Delta_j{a}^\dagger_ja_j+\omega_j{b}^\dagger_jb_j-g{a}^\dagger_ja_j({b}^\dagger_j+b_j)+iE({a}^\dagger_j-a_j)]\\
&-\mu(b_1{b}^\dagger_2+{b}^\dagger_1b_2)+\lambda(a_1{a}^\dagger_2+{a}^\dagger_1a_2)
\end{split}
\end{equation}
here ${a}^\dagger_j $  and $a_j $ are the optical creation and annihilation operators for the system $j$, ${b}^\dagger_j $ and $b_j $ are the mechanical creation and annihilation operators. $\Delta_j$ and $\omega_j$ are the optical detunings and the mechanical frequencies, respectively. $g$ is the optomechanical coupling constant and $E$ is the laser intensity which drives the optical cavities. $\mu$ is the intensity of the phonon tunnel and $\lambda$ is coupling constant of the fiber.The switches $K_1$ and $K_2$ can change $\mu$ and $\lambda$ values from zero to a positive constant by on and off. After considering the dissipative effects, the following quantum Langevin equations can be written in Heisenberg picture through the input-output properties
\begin{equation}
\begin{split}
&\partial_{t}a_1=[-\kappa+i\Delta_1+ig({b}^\dagger_1+b_1)]a_1+E-i\lambda a_2+\sqrt{2\kappa}{a}^{in}_1\\ 
&\partial_{t}a_2=[-\kappa+i\Delta_2+ig({b}^\dagger_2+b_2)]a_2+E-i\lambda a_1+\sqrt{2\kappa}{a}^{in}_2\\ 
&\partial_{t}b_1=[-\gamma-i\omega_1]b_1+ig{a}^\dagger_1a_1+i\mu b_2+\sqrt{2\gamma}{b}^{in}_1\\ 
&\partial_{t}b_2=[-\gamma-i\omega_2]b_2+ig{a}^\dagger_2a_2+i\mu b_1+\sqrt{2\gamma}{b}^{in}_2
\end{split}
\end{equation}
here $\kappa$ and $\gamma$ are the optical and mechanical damping rates. ${a}^{in}_j$ and ${b}^{in}_j$ are the input bath operators, which satisfy $\langle{a}^{in}_j{(t)}^{\dagger}{a}^{in}_{j'}{(t')}+{a}^{in}_{j'}{(t')}{a}^{in}_j{(t)}^{\dagger}\rangle=\delta_{jj'}\delta(t-t')$ and
$\langle{b}^{in}_j{(t)}^{\dagger}{b}^{in}_{j'}{(t')}+{b}^{in}_{j'}{(t')}{b}^{in}_j{(t)}^{\dagger}\rangle=(2n_b+1)\delta_{jj'}\delta(t-t')$, where $n_b=[\exp (\dfrac{\hbar\omega_j}{k_BT})-1]^{-1}$

Using the expectation value and quantum fluctuation to replace operators from Eq.(10), we can get following two equations.

The parts of expectation value are
\begin{equation}
\begin{split}
&\partial_{t}A_1=[-\kappa+i\Delta_1+ig({B}^*_1+B_1)]A_1+E-i\lambda A_2\\ 
&\partial_{t}A_2=[-\kappa+i\Delta_2+ig({B}^*_2+B_2)]A_2+E-i\lambda A_1\\ 
&\partial_{t}B_1=[-\gamma-i\omega_1]B_1+ig{A}^*_1A_1+i\mu B_2\\ 
&\partial_{t}B_2=[-\gamma-i\omega_2]B_2+ig{A}^*_2A_2+i\mu B_1
\end{split}
\end{equation}
where $A_j=\langle a_j\rangle$ and $B_j=\langle b_j\rangle$.

The parts of quantum fluctuation are
\begin{equation}
\begin{split}
&\partial_{t}\delta a_1=[-\kappa+i\Delta_1+ig({B}^*_1+B_1)]\delta a_1+igA_1({\delta b}^\dagger_1+\delta b_1)-i\lambda{\delta}a_2+\sqrt{2\kappa}{a}^{in}_1\\ 
&\partial_{t}\delta a_2=[-\kappa+i\Delta_2+ig({B}^*_2+B_2)]\delta a_2+igA_2({\delta b}^\dagger_2+\delta b_2)-i\lambda{\delta}a_1+\sqrt{2\kappa}{a}^{in}_2\\ 
&\partial_{t}\delta b_1=[-\gamma-i\omega_1]\delta b_1+ig{A}^*_1\delta a_1+igA_1{\delta a}^\dagger_1 +i\mu\delta b_2+\sqrt{2\gamma}{b}^{in}_1\\ 
&\partial_{t}\delta b_2=[-\gamma-i\omega_2]\delta b_2+ig{A}^*_2\delta a_2+igA_2{\delta a}^\dagger_2 +i\mu\delta b_1+\sqrt{2\gamma}{b}^{in}_2
\end{split}
\end{equation}

The dynamic properties of the cavity and oscillator can be described by their own conjugate mechanical quantities, i.e.
\begin{equation}
\begin{split}
&x_j=({a}^\dagger_j+a_j)/\sqrt{2}\\
&y_j=i({a}^\dagger_j-a_j)/\sqrt{2}\\
&q_j=({b}^\dagger_j+b_j)/\sqrt{2}\\
&p_j=i({b}^\dagger_j-b_j)/\sqrt{2}
\end{split}
\end{equation}

Substituting Eq.(13) into Eq.(12), Eq.(12) can be expressed in the matrix form
\begin{equation}
\partial_tu=Su+\xi
\end{equation}
where $u$ is a vector $(\delta x_1,\delta y_1,\delta x_2,\delta y_2,\delta q_1,\delta p_1,\delta q_2,\delta p_2)^{\top}$ and $\xi$ means a input vector \\
$(\delta {x}^{in}_1,\delta {y}^{in}_1,\delta {x}^{in}_2,\delta {y}^{in}_2,\delta {q}^{in}_1,\delta {p}^{in}_1,\delta {q}^{in}_2,\delta {p}^{in}_2)^{\top}$. $ S $ is $8\times8$ time-dependent matrix.
\begin{equation}
\begin{pmatrix}
 -\kappa & -\Delta_1-2g\text{Re}[B_1]& 0&  \lambda&  -2g\text{Im}[A_1]& 0&  0&0 \\ 
 \Delta_1+2g\text{Re}[B_1]&  -\kappa&  -\lambda&  0&  2g\text{Re}[A_1]&  0&  0&0 \\ 
 0&  \lambda& -\kappa & -\Delta_2-2g\text{Re}[B_2] &0&  0&  -2g\text{Im}[A_2]&  0 \\ 
 -\lambda&  0&  \Delta_2+2g\text{Re}[B_2]&  -\kappa&  0&  0&  2g\text{Re}[A_2]&0 \\ 
 0&  0&  0&  0&  -\gamma&  \omega_1&  0&-\mu \\ 
 2g\text{Re}[A_1]&  2g\text{Im}[A_1] &  0&  0&  -\omega_1&  -\gamma&  \mu& 0\\ 
 0&  0&  0&  0&  0&  -\mu&  -\gamma& \omega_2\\ 
 0&  0&  2g\text{Re}[A_2]&  2g\text{Im}[A_2]&  \mu&  0& -\omega_2 & -\gamma
\end{pmatrix}
\end{equation}

For the convenience of calculation, a covariance matrix $C$ is defined as
\begin{equation}
c_{ij}(t)=c_{ji}(t)=\dfrac{1}{2}\langle u_i(t)u_j(t)+u_j(t)u_i(t)\rangle
\end{equation}
and the evolution of matrix $C$ can be determined by Eq.(17):
\begin{equation}
\partial_{t}C=SC+CS^{\top}+N
\end{equation}
where $N$ is a diagonal noise correlation matrix defined by:
\begin{equation}
N_{ij}\delta(t-t')=\dfrac{1}{2}\langle \xi_i(t)\xi_j(t')+\xi_j(t')\xi_i(t)\rangle
\end{equation}

At this point, the dynamical analysis of the model we used has already been finished and the system will evolution according to Eq.(11) and Eq.(14). Subsequently, we are going to discuss the synchronization effects between the systems by using those equations.

We will discuss the phase synchronization between the oscillators of the systems. Firstly, define the ``classical" part (expectation value) of phase error on the basis of Eq.(6) [22].
\begin{equation}
\langle \phi_-(t)\rangle=[\langle \phi_1(t)\rangle-\langle \phi_2(t)\rangle]/\sqrt{2}=\theta(t)/\sqrt{2}
\end{equation}
where
\begin{equation}
\begin{split}
\langle \phi_1(t)\rangle=\text{arg}[B1(t)]=\arctan (\text{Im}[B1(t)]/\text{Re}[B1(t)])\\
\langle \phi_2(t)\rangle=\text{arg}[B2(t)]=\arctan (\text{Im}[B2(t)]/\text{Re}[B2(t)])
\end{split}
\end{equation}

The evolution of $\theta(t)$ can be simulated numerically with simultaneous equations (11), (19) and (20) under the certain initial conditions and the largest Lyapunov exponent of the error can be calculated by following equation
\begin{equation}
L_y=\lim_{t\rightarrow\infty}\dfrac{1}{t}\ln\left|\dfrac{\delta\theta(t)}{\delta\theta(0)}\right|
\end{equation}

$\delta\theta(0)$ and $\delta\theta(t)$ in Eq.21 mean the disturbances of the phase errors when $t=0$ and $t=t$, respectively, and $L_y$ can be seen as the eigenvalue of Jacobian matrix corresponding to $\delta\theta$. Because the phase synchronization is controlled by the intensity of the phonon tunnel and the coupling constant of the fiber, we calculate the largest Lyapunov exponent of phase error with the variation of $\mu$ and $\lambda$. In calculation, the damping rates and the intensity of driving field are assumed to be equal in both systems, but there are the differences in frequencies and initial conditions. The values of $\omega$,$g$,$\kappa$ and $\gamma$ are taken as the same as Mari's work so that the conclusion is more easy to be verified by the experiment. Moreover, we properly reduce the intensity of the driving field in order to highlight the coupling function in synchronization. Otherwise, too strong driving field will dilute the coupling effect, which leads the system in the ``forced" synchronous effect. It can be known from Eq.9 and Eq.11 that the phonon coupling can directly influence mechanical oscillators, however, the photon coupling can only influence them indirectly by changing the light field in the cavities. For making two switches have similar ability to control synchronization, we reduce the intensity of phonon channel($\mu$) and increase coupling constant of the fiber. Therefore, we calculate the Lyapunov exponent in the region of $\lambda\in[0,0.2]$, $\mu\in[0,0.01]$ and the calculation result is shown in Fig.2.

\begin{figure}[h!]
\centering
\includegraphics[scale=0.37]{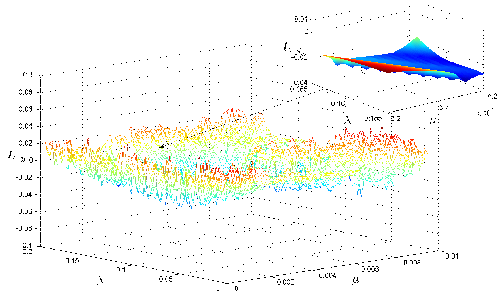}
\caption{Largest Lyapunov exponent of phase error $\theta(t)$ with $\mu$ and $\lambda$. Here $ \omega_1=1$ is the unit of frequency, and $ \omega_2=1.005$, $ g=0.004$, $\kappa=0.15$, $\gamma=0.005$, $\Delta_j=\omega_j$, $n_b=0$.}
\label{threadsVsSync}
\end{figure}

Instead of analyzing the largest Lyapunov exponent concretely, as it is discussed above, we can determine whether the system is in the synchronization state only by comparing the largest Lyapunov exponent with zero. Therefore, Fig.2 is redrawn in this form: some parameter regions where the largest Lyapunov exponent is greater than zero are projected and marked in red, contrary, the regions where the largest Lyapunov exponent is less than zero are marked in blue. Moreover, we also plot a curve of the Lyapunov exponent with a fixed $\lambda$ in order to display it more clearly, shown in Fig.3.

\begin{figure}[!h]
\centering
\subfigure[]{    
\label{fig:subfig:a}   
\includegraphics[scale=0.5]{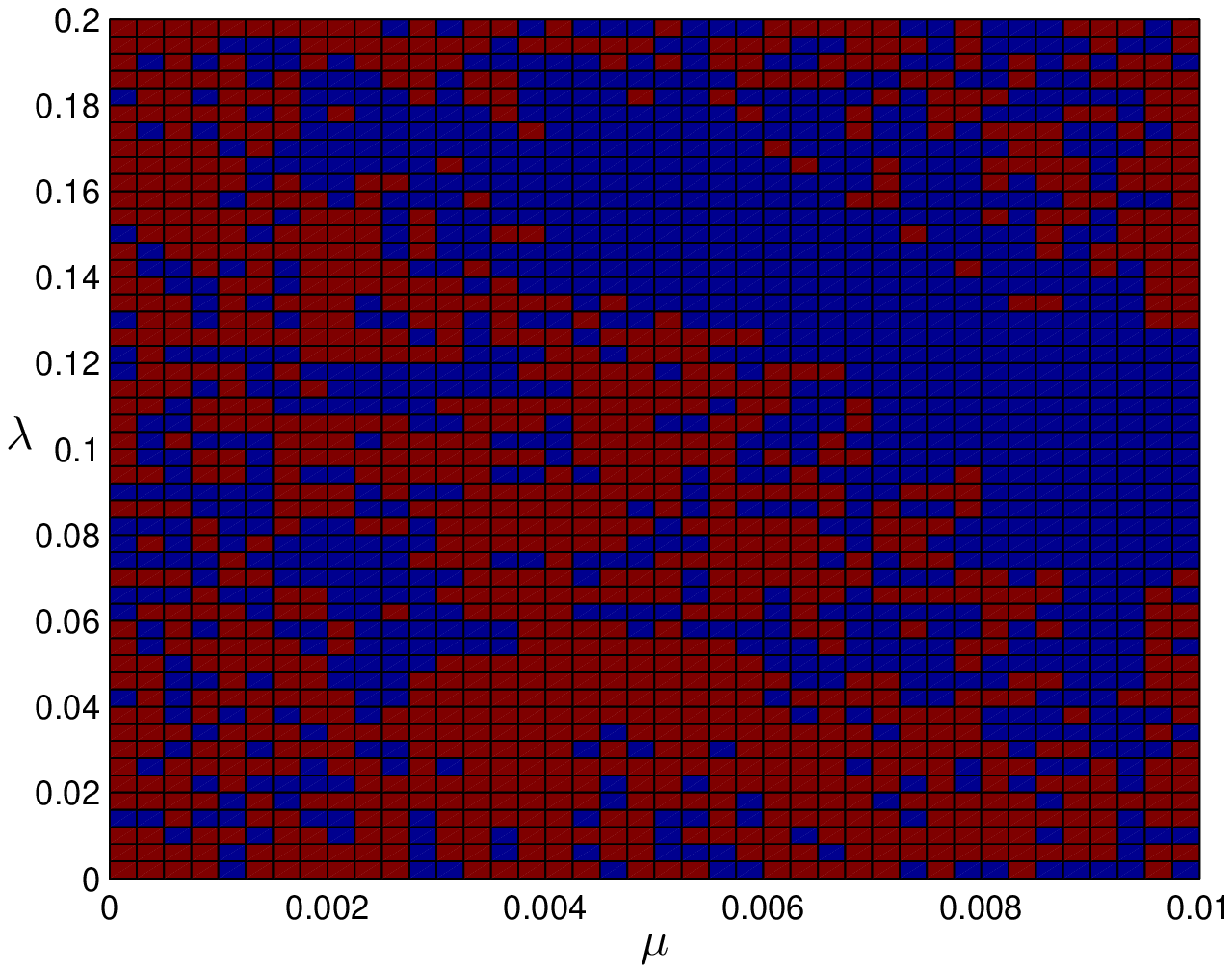}}  
\hspace{0in}  
\subfigure[]{    
\label{fig:subfig:b}   
\includegraphics[scale=0.5]{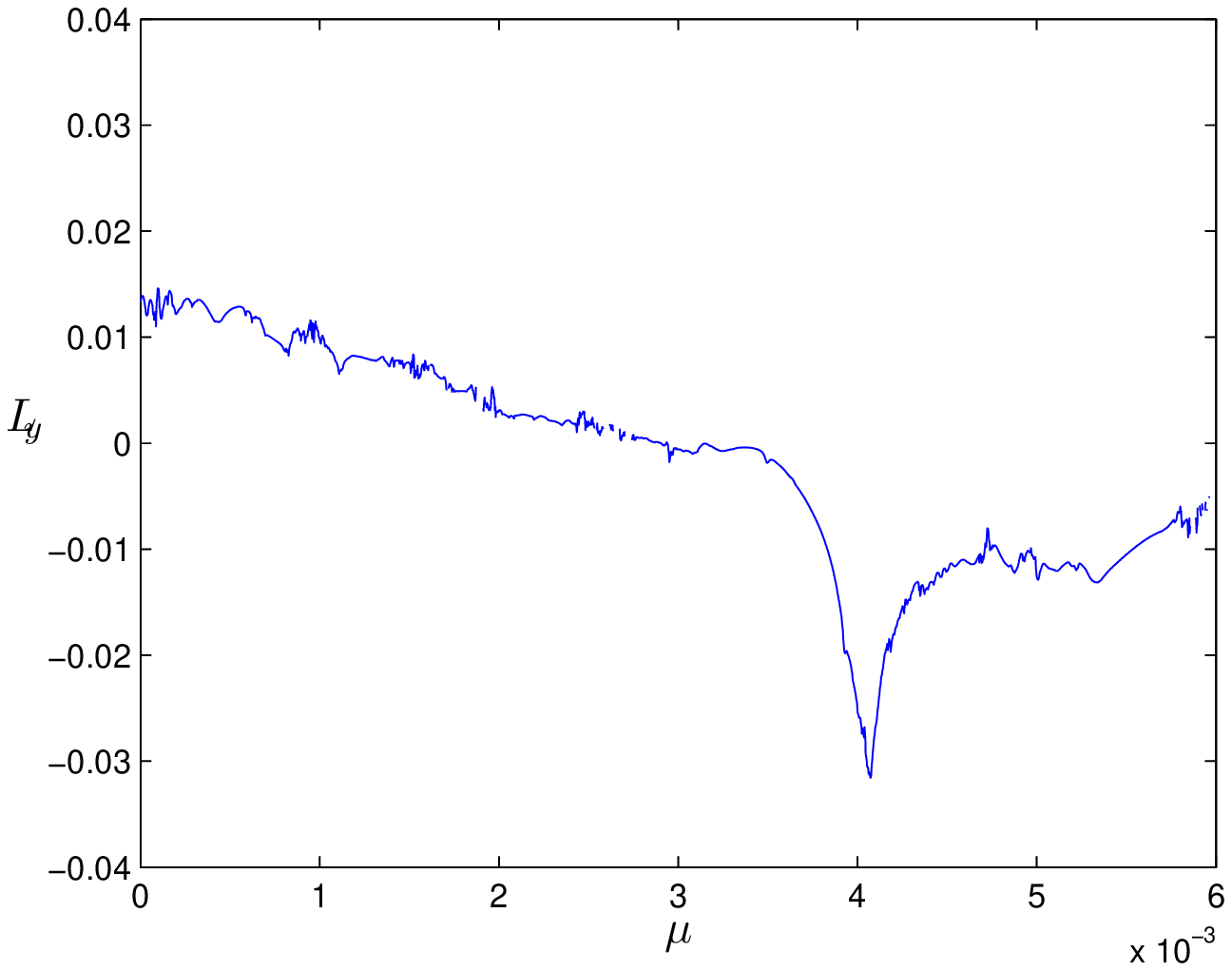}}    
\caption{ (a): Comparison between the largest Lyapunov exponent and zero. Blue areas mean the largest Lyapunov exponent is less than zero and red areas mean the largest Lyapunov exponent is greater than zero. (b): Evolution of the largest Lyapunov exponent with varied $\mu$. Here we set $\lambda=0.16$ and other parameters are same with Fig.2.
\label{fig:fig1a}}
\end{figure}

If $\mu$ and $\lambda$ are in the blue regions of Fig.3(a), the largest Lyapunov exponent is less than zero, indicating the evolution of the phase error tends to zero stably after a certain time and the systems reach the synchronization. By contrast, the largest Lyapunov exponent is greater than zero while $\mu$ and $\lambda$ are in the red regions and the systems are not synchronous because the phase error tends to random oscillations. Therefore, it can be seen from Fig.3(a) that the systems will not synchronize when two systems are not coupled($\mu=0$ and $\lambda$=0). Once there is coupling between systems, however, the red regions will be replaced gradually by blue area with the increasing  of $\mu$ and $\lambda$. We expect to control the synchronization by the switches $K_1$ and $K_2$ together, i.e. it will be happened only if two switches  meet the logic ``AND". Then we choose the parameters according to the following principles: the largest Lyapunov exponent is greater than zero when $\mu=0$ and $\lambda\neq 0$ as well as $\mu\neq 0$ and $\lambda= 0$, but it must be less than zero as $\mu\neq 0$ and  $\lambda\neq 0$ at the same time. According to this, the appropriate parameters can be found in $\lambda\in[0.14,0.2]$ and $\mu\in[0.004,0.007]$ recurring to the help of Fig.3 and within the range the systems will achieve phase synchronization only if switches $K_1$ and $K_2$ are closed synchronously. (It is worth to note that alone point with different color from ones around can be ignored as an error.) In other words, we can control synchronization effect with switches  $K_1$ and $K_2$ by using above characteristics.

In the above discussion, we give a range of parameters $\mu$ and $\lambda$ but not exact values. This is because the ``classical'' error can tend to zero for all parameters in that range. Whereas for quantum synchronization, not only the expectation value of errors tending to zero is necessary, but also the error fluctuations as small as possible. For the sake of reaching a perfect synchronization, we need to select appropriate values of $\mu$ and $\lambda$ and to ensure a minimum quantum fluctuations. It is the reason why we calculate the second order measurement $S_p'(t)$ of the quantum phase synchronization.

For the calculating of $S_p'(t)$ , the matrix $C$ defined in Eq.(16) is transformed firstly as
\begin{equation}
C'(t)=U(t)C(t)U(t)^{\dagger}
\end{equation}
here $U(t)=diag[e^{-i\phi_{a1}t},e^{i\phi_{a1}t},...]$ and $\phi_{a1}=\text{arg}\langle a_1(t)\rangle$, $\phi_{a2}=\text{arg}\langle a_2(t)\rangle,...$ . $C'(t)$ can be obtained by substituting the matrix $C'$ into Eq.(17) and $ S_p'(t)$ can be expressed as
\begin{equation}
\begin{split}
S_p'(t)={\dfrac{1}{2}\langle{\delta\phi_-(t)}^{2}\rangle}^{-1}&=\dfrac{1}{2}{\langle\dfrac{1}{2}(\delta {p'}_1^2+\delta {p'}_2^2-2\delta {p'}_1\delta {p'}_2)\rangle}^{-1}\\
&=\dfrac{1}{2}{[\dfrac{1}{2}(C'_{66}+C'_{88}-2C'_{68})]}^{-1}
\end{split}
\end{equation}

Time-averaged $S_p'(t)$ is further calculated in order to show directly the size of quantum fluctuation under the different parameters.

\begin{equation}
\overline{S_p'}=\lim_{T\rightarrow\infty}\dfrac{1}{T}\int_{0}^{T}S_p'(t)dt
\end{equation}
and the calculation result is shown in Fig.4.

\setcounter{figure}{3}
\renewcommand\thefigure{\arabic{figure}}
\begin{figure}[h!]
\centering
\includegraphics[scale=0.45]{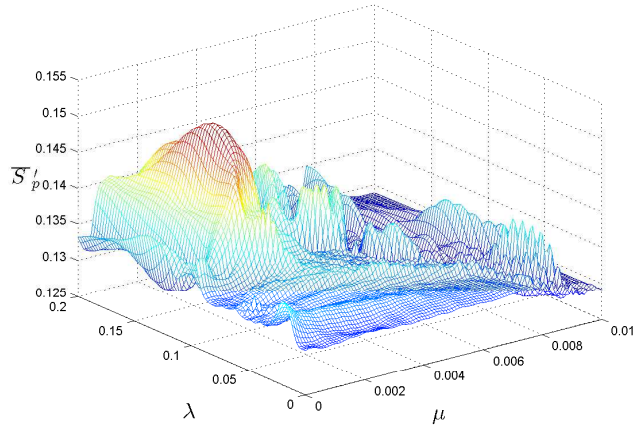}
\caption{$\overline{S_p'}$ with varied $\mu$ and $\lambda$. In this calculation, we let $T=2000$ and other parameters are same with Fig.2}
\label{threadsVsSync}
\end{figure}

Although the phase expectation values of every system tend to be equal, the quantum fluctuations between the systems under different parameters still influence the perfection of quantum phase synchronization, as shown in Fig.4. The fluctuation of system error will be reduced to minimum extent while taking $\mu=0.004$ and $\lambda=0.16$, which draws the conclusion that the best effect of synchronization has been reached.

The dynamical evolution of the system is simulated here to test the validity of our criterion. Before the simulation we let $\mu=0.004$ and $\lambda=0.16$ and assume that the initial phase error between the systems is $\theta(0)=\dfrac{\pi}{2}$. The remaining parameters are same as ones used in Fig.2. The simulation results are illustrated in Fig.5 in which the unit of ordinate is $\pi$.

\begin{figure}[!h]
\centering
\subfigure[]{    
\label{fig:subfig:a}   
\includegraphics[scale=0.24]{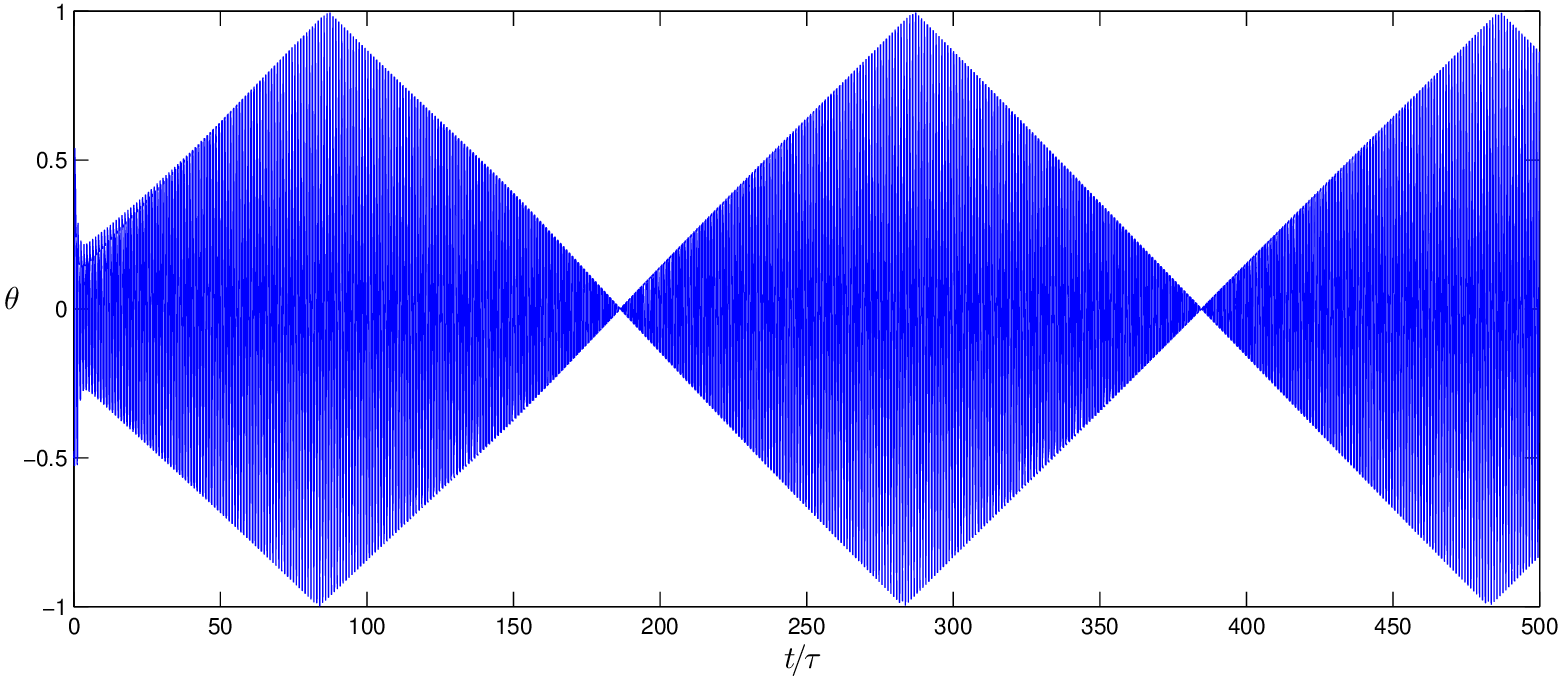}}  
\hspace{0in}  
\subfigure[]{    
\label{fig:subfig:b}   
\includegraphics[scale=0.24]{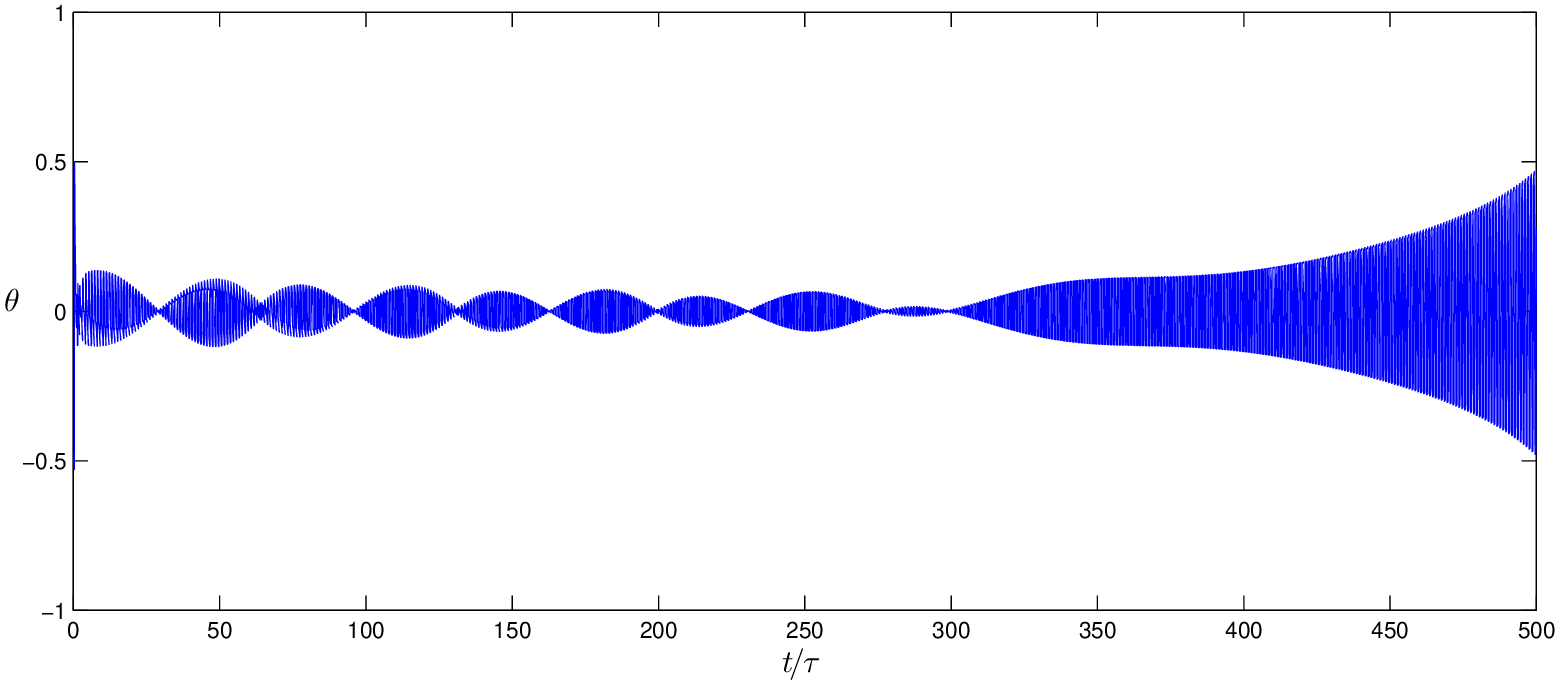}}   
\subfigure[]{    
\label{fig:subfig:c}   
\includegraphics[scale=0.24]{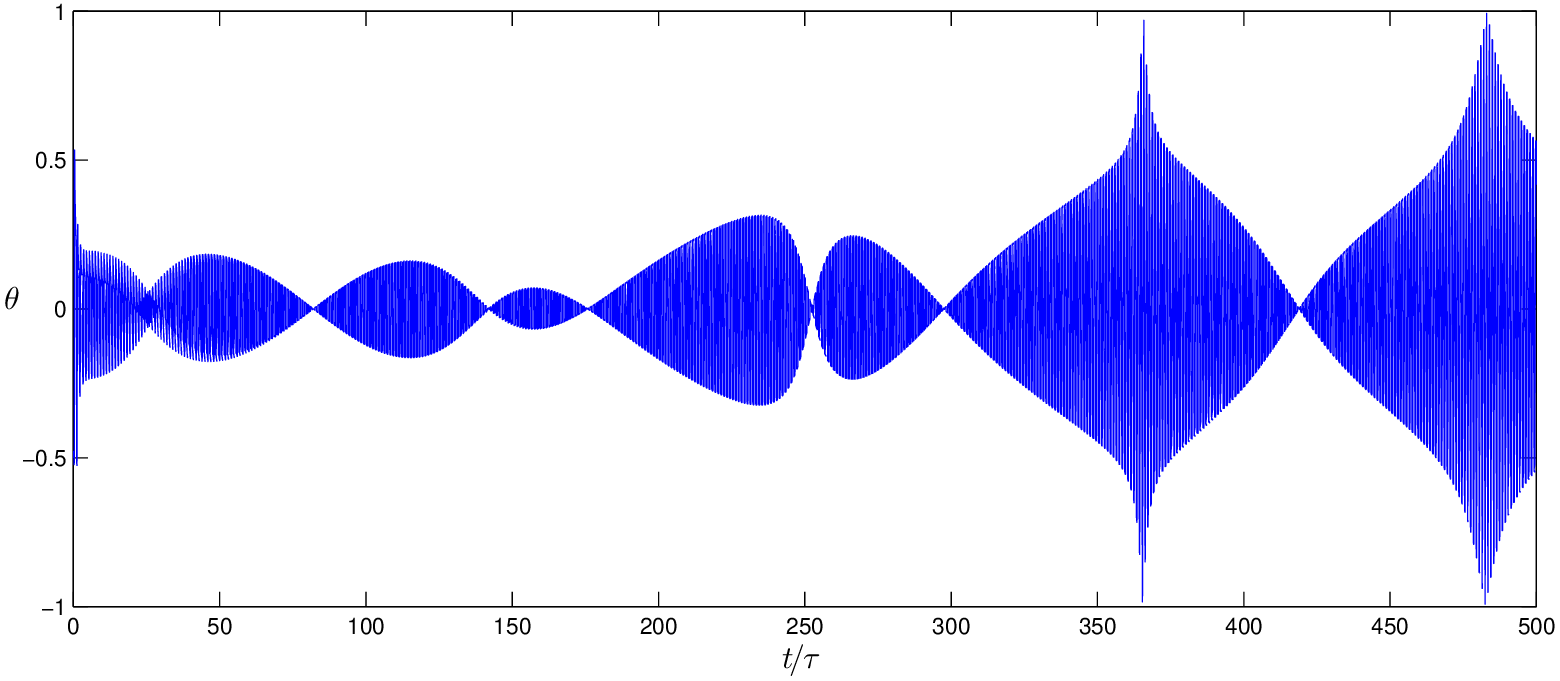}}  
\subfigure[]{    
\label{fig:subfig:d}   
\includegraphics[scale=0.24]{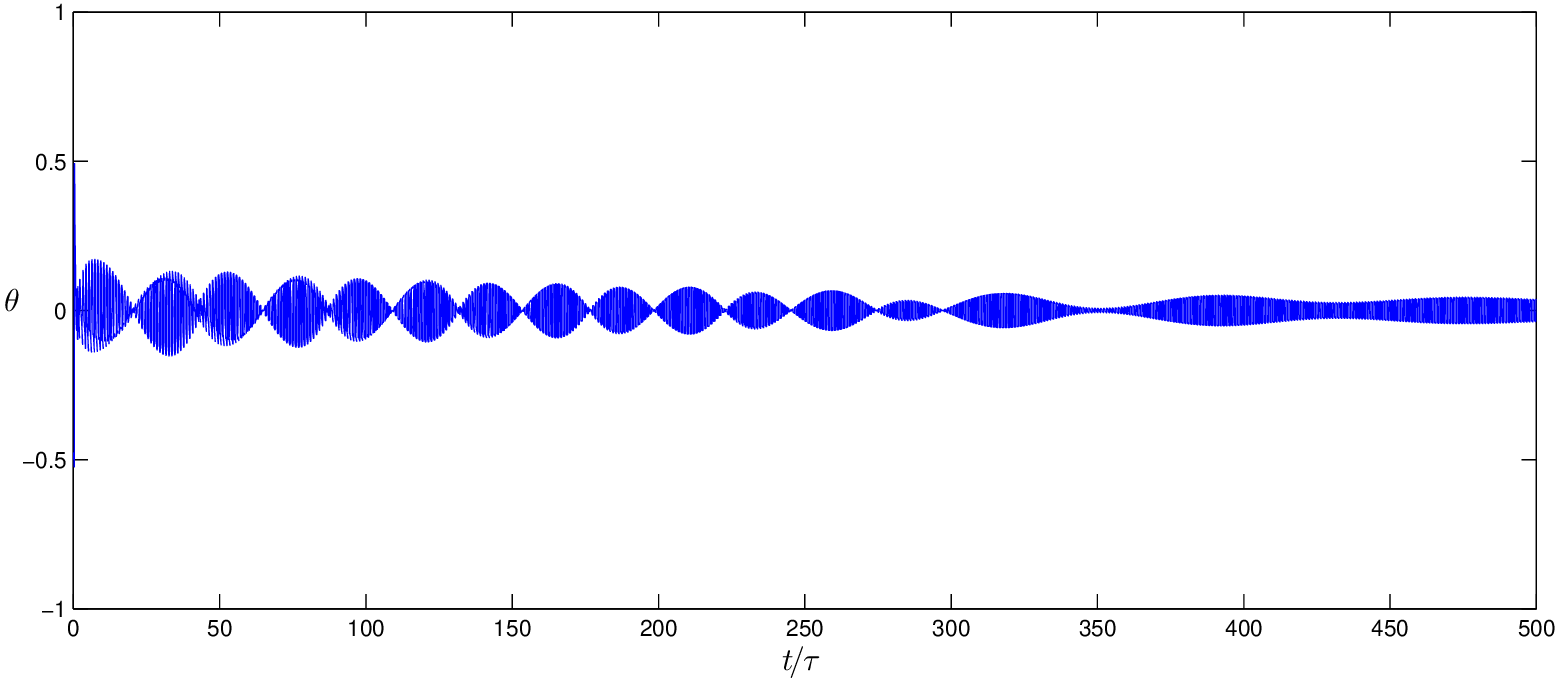}}    
\caption{Evolution of the phase error between the systems, (a): $K_1$ and $K_2$ are both opened ($\mu=0$,$\lambda=0$); (b): $K_1$ is opened and $K_2$ closed ($\mu=0$,$\lambda=0.16$); (c): $K_1$ is closed and $K_2$ opened ($\mu=0.004$,$\lambda=0$) and (d): $K_1$ and $K_2$ are both closed ($\mu=0.004$,$\lambda=0.16$).
\label{fig:fig1a}}
\end{figure}

It can be known from Figs.5(a)$\sim$(c) that the synchronization between systems will not be achieved as long as any a switch of $K_1$ and $K_2$ is opened. Upon further inspection, we notice that two different systems will never achieve phase synchronization in the other parameters ($\omega$,$g$,$\kappa$ and $\gamma$) but the same $E$ when couples disappear. Only when two switches are both closed, the systems can realize synchronization, shown as Fig.5(d). This result is identical with our analysis and it also verifies the quantum synchronous criterion proposed in our work. By the way, the logical relationship ``AND" of two switches is taken as an example in Fig.3, however, the logical relationships ``OR" or ``exclusive-OR" between the switches can also be selected to realize the quantum synchronization of the systems by adjusting appropriate parameters of $\mu$ and $\lambda $ .

\section{Conclusions}  

In this paper, we investigate quantum synchronization effects and present the quantitative criteria of complete synchronization and phase synchronization between quantum systems. Further, we realize the quantum phase synchronization between coupled optomechanical systems by using our criterisa. Through calculating the largest Lyapunov exponent, we find that the systems will not reach synchronization unless switches $K_1$ and $K_2$ are closed synchronously (satisfying the logic relation ``AND") when the parameter values are taken at the ranges $\lambda\in[0.14,0.2]$ and $\mu\in[0.004,0.007]$. At the same time, we obtain the information that the fluctuation of system error will reduce to minimum while
 $\mu=0.004$ and $\lambda=0.16$ by calculating the second order measurement $S_p'(t) $ of the quantum phase synchronization. Finally, the dynamical evolution of the system is simulated in order to test the validity of our criterion under above parameters. Since the concrete quantum synchronization criteria have been proposed and the control theory of quantum synchronization effects is simple and efficient in the work, other designers can set different synchronization conditions to satisfy themselves aims. We believe that our work can bring certain application values in quantum communication, quantum control and quantum logical gates.
\subsection*{Acknowledgements}
This research was supported by the National Natural Science Foundation of China (Grant No 11175033) and the Fundamental Research Funds for the Central Universities (DUT13LK05).


\subsection*{References}
\setlength{\parindent}{0pt}

[1]Yamada T, Fujisaka H. Stability theory of synchronized motion in coupled- oscillator systems \uppercase\expandafter{\romannumeral2}. Prog. Theor. Phys., 1983, 70(5):1240-1248.

[2]Pecora L M, Carroll T L. Synchronization in chaotic systems. Phys. Rev. Lett., 1990, 64(8):821-824.

[3]Selivanov A A, Lehnert J, Dahms T, Hövel P, Fradkov A L, Schöll E. Adaptive synchronization in delay-coupled networks of Stuart-Landau oscillators. Phys. Rev. E, 2012, 85(1):016201-8.

[4]Fürthauer S, Ramaswamy S. Phase-synchronized state of oriented active fluids. Phys. Rev. Lett., 2013, 111(23):238102-5.

[5]Paredes G, Alvarez-Llamoza O, Cosenza M G. Global interactions, information flow, and chaos synchronization. Phys. Rev. E, 2013, 88(4):042920-8.

[6]Um J, Hong H, Park H. Nature of synchronization transitions in random networks of coupled oscillators. Phys. Rev. E, 2014, 89(1):012810-8.

[7]Kawamura Y, Nakao H. Noise-induced synchronization of oscillatory convection and its optimization. Phys. Rev. E, 2014, 89(1):012912-13.

[8]Kippenberg T J, Vahala K J. Cavity optomechanics: back-action at the mesoscale. Science, 2008, 321(5893): 1172-1176.

[9]Thompson J D, Zwickl B M, Jayich A M, Marquardt F, Girvin S M,Harris J G E. Strong dispersive coupling of a high-finesse cavity to a micromechanical membrane. Nature, 2008, 452(7183): 72-75.

[10]Brennecke F, Donner T, Ritter S, Bourdel T, Köhl M, Esslinger T. Cavity QED with a Bose–Einstein condensate. Nature, 2007, 450(7167): 268-271.

[11]Lin Q, Rosenberg J, Jiang X, K J Vahala, O Painter. Mechanical oscillation and cooling actuated by the optical gradient force. Phys. Rev. Lett., 2009, 103(10): 103601-4.

[12]Zhang K, Chen W, Bhattacharya M, Meystre P. Hamiltonian chaos in a coupled BEC–optomechanical-cavity system. Phys. Rev. A, 2010, 81(1): 013802-6.

[13]Larson J, Horsdal M. Photonic Josephson effect, phase transitions, and chaos in optomechanical systems. Phys. Rev. A, 2011, 84(2): 021804-4.

[14]Heinrich G, Harris J G E, Marquardt F. Photon shuttle: Landau-Zener-Stückelberg dynamics in an optomechanical system. Phys. Rev. A, 2010, 81(1): 011801-4.

[15]Vitali D, Gigan S, Ferreira A, Böhm H R, Tombesi P, Guerreiro A, Vedral V, Zeilinger A, Aspelmeyer M. Optomechanical entanglement between a movable mirror and a cavity field. Phys. Rev. Lett., 2007, 98(3): 030405-4.

[16]Miao H, Danilishin S, Chen Y. Universal quantum entanglement between an oscillator and continuous fields. Phys. Rev. A, 2010, 81(5): 052307-4.

[17]Hofer S G, Vasilyev D V, Aspelmeyer M, Hammerer K. Time-Continuous Bell Measurements. Phys. Rev. Lett., 2013, 111(17): 170404-6.

[18]Wang Y D, Clerk A A. Reservoir-engineered entanglement in optomechanical systems. Phys. Rev .Lett., 2013, 110(25): 253601-5.

[19]Mari A, Farace A, Didier N, Giovannetti V, Fazio R. Measures of quantum synchronization in continuous variable systems. Phys. Rev. Lett., 2013, 111(10): 103605-5.

[20]Farace A, Giovannetti V. Enhancing quantum effects via periodic modulations in optomechanical systems. Phys. Rev. A, 2012, 86(1): 013820-12.

[21]Mari A, Eisert J. Gently modulating optomechanical systems. Phys. Rev. Lett., 2009, 103(21): 213603-4.

[22]In Eq.(2) and Eq.(6), we make the error operators divided by $\sqrt{2}$ artificially in order to ensure $0<S_c(t)\leq1$.Discussing the expectation value of the error operator, however, the physical significance of error will not clear if make the error operators divided by $\sqrt{2}$. So, we substitute $\theta(t)=\langle\phi_1(t)\rangle-\langle\phi_2(t)\rangle$ into Eq.(21), instead of $ \langle\phi_-(t)\rangle=(\langle\phi_1(t)\rangle-\langle\phi_2(t)\rangle)/\sqrt{2}$.

[23]Lee T E ,Chan C-K ,Wang S S. Entanglement tongue and quantum synchronization of disordered oscillators. Phys. Rev. E, 2014, 89(2): 022913-10

[24]Matheny M H, Grau M, Villanueva L G, Karabalin R B, Cross M C, Roukes M L. Phase Synchronization of Two Anharmonic Nanomechanical Oscillators. Phys. Rev. Lett., 2014, 112(1): 014101-5

[25]Manipatruni S,Weiderhecker G,Lipson M. Long-range synchronization of optomechanical structures. Quantum Electronics and Laser Science Conference. Optical Society of America, 2011: QWI1

[26]Aspelmeyer M, Kippenberg T J, Marquardt F. Cavity optomechanics, arXiv:1303.0733v1

[27]Heinrich G,Ludwig M, Qian J, Kubala B, Marquardt F. Phys. Rev. Lett., 2011,107(4):043603-4

\end{document}